\documentclass[12pt, letterpaper]{article}
\usepackage[titletoc,title]{appendix}
\usepackage{color}
\usepackage{booktabs}
\usepackage[usenames,dvipsnames,svgnames,table]{xcolor}
\definecolor{dark-red}{rgb}{0.75,0.10,0.10}
\definecolor{bluish}{rgb}{0.05,0.05,0.85}

\usepackage[margin=1in]{geometry}
\usepackage[linkcolor=blue,
			colorlinks=true,
			urlcolor=blue,
			pdfstartview={XYZ null null 1.00},
			pdfpagemode=UseNone,
			citecolor={bluish},
			pdftitle={Street Sense}]{hyperref}

\usepackage{natbib}
\usepackage{float}

\usepackage{geometry} % see geometry.pdf on how to lay out the page. There's lots.
\usepackage{graphicx}                % Handles inclusion of major graphics formats and allows use of 
\usepackage{amsfonts,amssymb,amsbsy}
\usepackage{amsxtra}
\usepackage{verbatim}
\setcitestyle{round,semicolon,aysep={},yysep={;}}
\usepackage{setspace}		     % Permits line spacing control. Options are \doublespacing, \onehalfspace
\usepackage{sectsty}		     % Permits control of section header styles
\usepackage{pdflscape}
\usepackage{fancyhdr}		     % Permits header customization. See header section below.
\usepackage{url}                     % Correctly formats URLs with the \url{} tag
\usepackage{fullpage}		%1-inch margins
\usepackage{multirow}
\usepackage{rotating}
\usepackage[T1]{fontenc}
\usepackage[bitstream-charter]{mathdesign}

\usepackage{chngcntr}
\usepackage{booktabs}
\usepackage{longtable}

\usepackage{afterpage}

\makeatother

\usepackage[hang, font=small,skip=0pt]{caption}
\usepackage{subcaption}

\makeatletter
\providecommand\phantomcaption{\caption@refstepcounter\@captype}
\makeatother

\title{Street Sense: Learning from Google Street View\footnote{Data and scripts behind the analysis presented here can be downloaded from \url{https://github.com/geosensing/streetsense}.
}}

\author{Suriyan Laohaprapanon\thanks{Suriyan can be reached at: \href{mailto:suriyant@gmail.com}{\texttt{suriyant@gmail.com}}} \and Kimberly Ortleb\thanks{Kimberly can be reached at: \href{kim.ortleb@gmail.com}{\texttt{kim.ortleb@gmail.com}}} \and Gaurav Sood\thanks{Gaurav can be reached at: \href{gsood07@gmail.com}{\texttt{gsood07@gmail.com}}}}

\begin{document}
\maketitle
\thispagestyle{empty}

\begin{abstract}
How good are the public services and the public infrastructure? Does their quality vary by income? These are vital questions---they shed light on how well the government is doing its job, the consequences of disparities in local funding, etc. But there is little good data on many of these questions. We fill this gap by describing a scalable method of getting data on one crucial piece of public infrastructure: roads. We assess the quality of roads and sidewalks by exploiting data from Google Street View. We randomly sample locations on major roads, query Google Street View images for those locations and code the images using Amazon's Mechanical Turk. We apply this method to assess the quality of roads in Bangkok, Jakarta, Lagos, and Wayne County, Michigan. Jakarta's roads have nearly four times the potholes than roads of any other city. Surprisingly, the proportion of road segments with potholes in Bangkok, Lagos, and Wayne is about the same, between .06 and .07. Using the data, we also estimate the relation between the condition of the roads and local income in Wayne, MI. We find that roads in more affluent census tracts have somewhat fewer potholes.
\end{abstract}

\clearpage

\doublespacing

The poorer the quality of the public infrastructure and public services, generally, the worse the quality of life. For instance, potholed roads mean that vehicles can't go as fast and the ride is bumpier. If sidewalks aren't paved, physically disabled have a tough time getting anywhere. If garbage isn't picked up regularly, foul smells and unsightliness are part of life, and the risk of disease is greater.  

As these examples convey, the quality of public infrastructure and public services matters immensely. It sheds light on the quality of life, and on the resources and functioning of the government. So how good is the public infrastructure? And how good are the public services? More often than not, we have no good answer to these questions.

In this paper, we introduce a method to answer questions about the quality of one important piece of public infrastructure: roads. We capitalize on \href{https://www.google.com/streetview/}{Google Street View} to learn about the condition of the roads. We randomly sample locations on the roads, get Google Street View images for those locations, and crowdsource the coding of the images. To illustrate the method's utility, we apply the method to learn about the condition of roads in Wayne (Michigan), Bangkok, Lagos, and Jakarta, and to assess the association between local income and the condition of the roads in Wayne. We also discuss ways this labeled data can be augmented and used to build automated systems to answer these questions at scale.

\section*{Learning From Google Street View}

Since 2007, Google has been working on regularly taking panoramic images of all the streets in the world. In the West, Google's efforts have been a success: Google's specially designed vehicles have traversed an overwhelming majority of the streets.\footnote{See \href{https://en.wikipedia.org/wiki/Coverage\_of\_Google\_Street\_View}{https://en.wikipedia.org/wiki/Coverage\_of\_Google\_Street\_View}} In the third-world, however, the coverage is patchy. For instance, as we show below, just about 24.6\% of Dhaka's streets are covered by Google Street View.\footnote{Some of Google's estimates of its coverage are either wrong or have become outdated as the road network continues to grow.} But Google's coverage of some other big third-world cities isn't too shabby. For instance, it covers 99.9\% of Bangkok's streets and 87.2\% of Lagos' streets. In all, the coverage is good enough, especially in the West, that people can build a scalable measurement infrastructure on top of it.

But patchy coverage is not the only problem with Google Street View data. The other is that the data are not always current. A large chunk of the data is at least a few years old. But somewhat older data has its value, especially because we expect Google to map those areas again in the future. The data aren't perfect but they are rich and valuable. 

But how do we efficiently capitalize on Google Street View data? We could download all the data for a city. But doing so is \href{https://developers.google.com/maps/documentation/streetview/usage-and-billing}{expensive}. And it may not even be useful. Depending on the question, a large random sample can fill in nicely for a census. For learning the condition of the roads, that is precisely the case.

To efficiently learn about the condition of the streets, sidewalks, and such, from Google Street View data, we devise a new workflow. We start by downloading data on the kinds of roads we are interested from \href{https://www.openstreetmap.org}{Open Street Map} (OSM). We then chunk the roads into half a kilometer segments, and then randomly sample from the segments. (The open source Python package \href{https://github.com/geosensing/geo_sampling}{geo-sampling} \citep{laoha2017} implements this workflow.) We then take the starting latitude and longitude of the sampled segments and query the \href{https://developers.google.com/maps/documentation/streetview/intro}{Google Street View API}.

\section*{Application}

To illustrate the utility of the method, we apply it to learn the condition of the roads, the condition of the sidewalks, and the presence of litter on the streets in four prominent third-world cities and one poor American county.

\begin{figure}[H]
\centering
\caption{Sampled Locations in Bangkok} \label{fig:google_bangkok}
    \includegraphics[width=\textwidth]{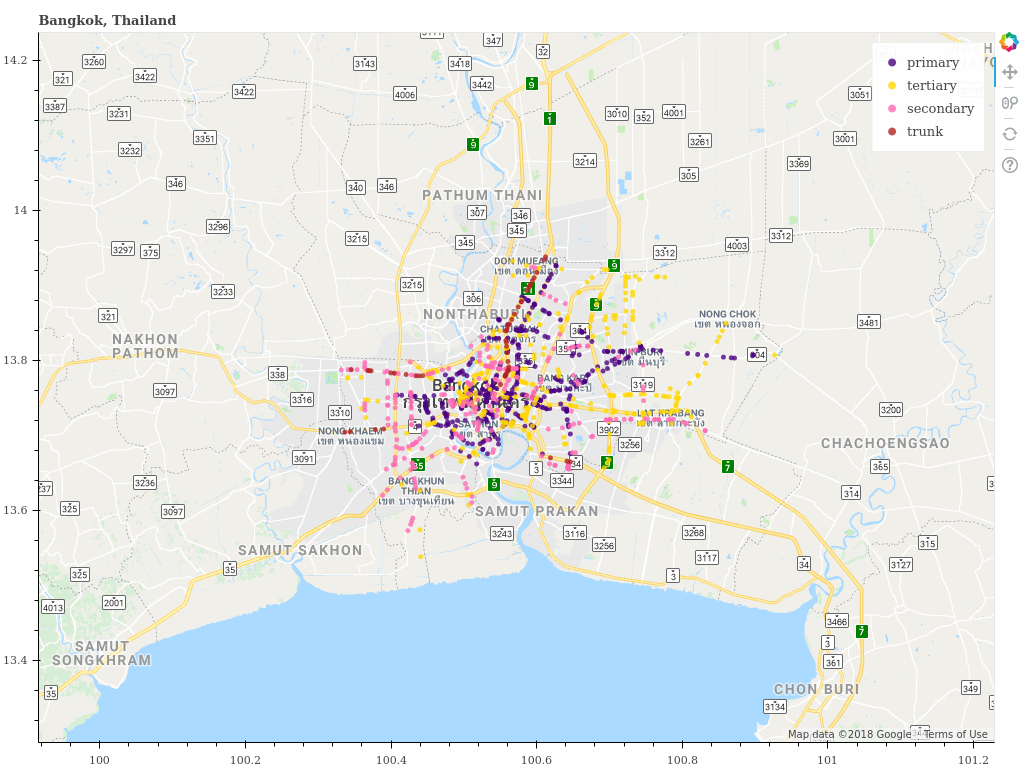}
\end{figure}

\begin{figure}[H]
\centering
\caption{Sampled Locations in Jakarta} \label{fig:google_jakarta}
    \includegraphics[width=\textwidth]{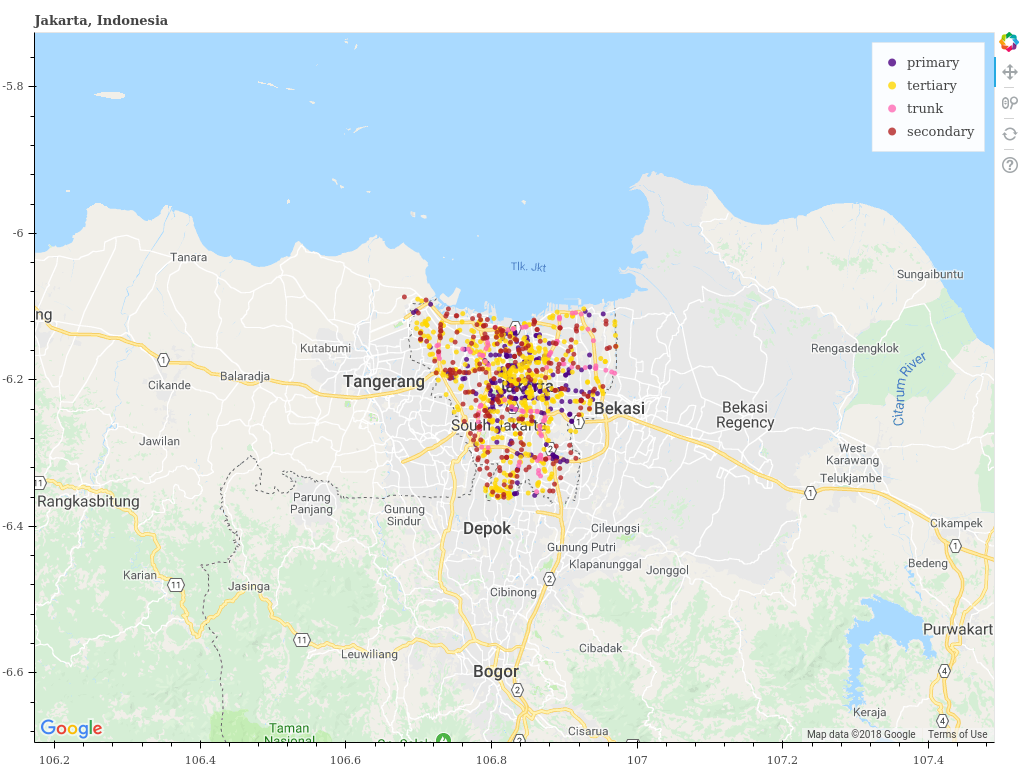}
\end{figure}

\begin{figure}[H]
\centering
\caption{Sampled Locations in Lagos} \label{fig:google_lagos}
    \includegraphics[width=\textwidth]{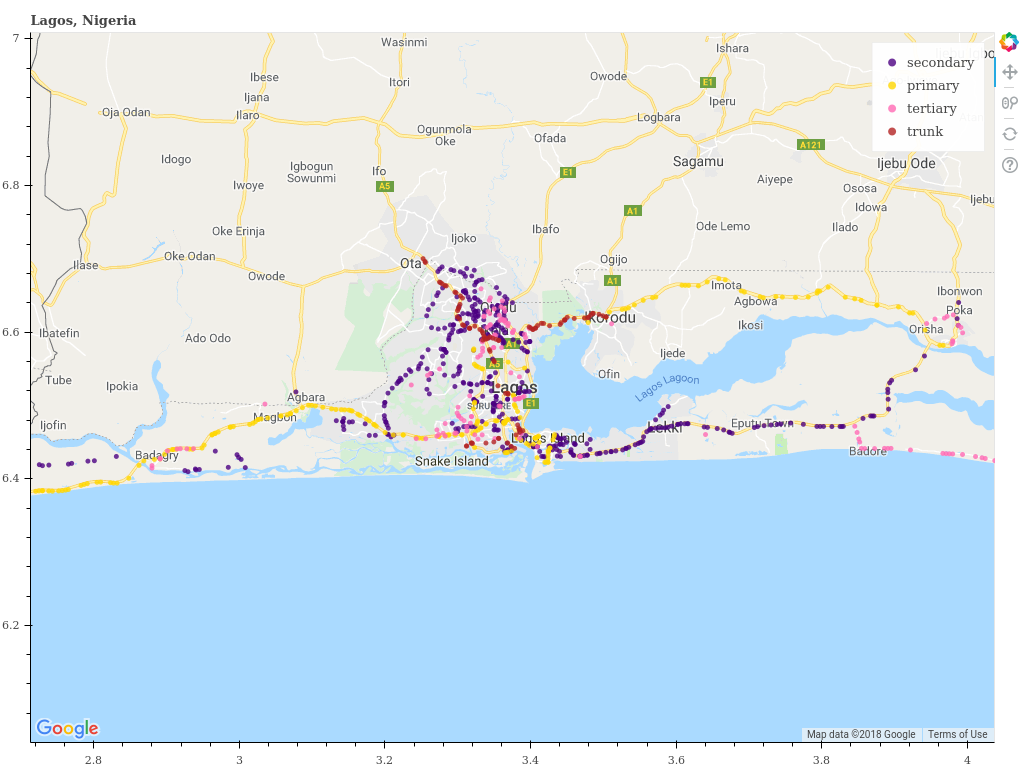}
\end{figure}

To learn the condition of roads in Bangkok, Dhaka, Jakarta, Lagos, and Wayne, MI, in the latter half of 2017, we downloaded data on all the streets from OSM. We feared that in many of these cities, Google Street View's coverage of neighborhood roads would be patchy. So we decided to focus on \href{https://wiki.openstreetmap.org/wiki/Tag:highway=primary}{primary}, \href{https://wiki.openstreetmap.org/wiki/Tag:highway=secondary}{secondary}, \href{https://wiki.openstreetmap.org/wiki/Tag:highway=tertiary}{tertiary}, and \href{https://wiki.openstreetmap.org/wiki/Tag:highway=trunk}{trunk} roads. We used the \href{https://github.com/geosensing/geo_sampling}{geo-sampling package} to take a random sample of primary, secondary, tertiary, and trunk road segments for each location (see Figures \ref{fig:google_bangkok}, \ref{fig:google_jakarta}, \ref{fig:google_lagos}). (Figures \ref{fig:bangkok_sample}, \ref{fig:jakarta_sample}, \ref{fig:lagos_sample}, \ref{fig:wayne2_sample} plot the starting longitude and latitude without the surrounding detail of the sampled segments of Bangkok, Jakarta, Lagos, and Wayne, MI respectively.) For Bangkok, Dhaka, Jakarta, and Lagos,  we drew a sample of 1,000 segments each. For Wayne, MI, we drew a sample of 5,000 segments. We drew a larger sample for Wayne, MI because we wanted to estimate the relationship between local income and road conditions there. (We chose an American county to estimate the relationship between local income and road conditions because data on local income is readily available for the US.)

\begin{figure}[H]
\centering
\caption{Sampled Locations in Wayne} \label{google_wayne2}
    \includegraphics[width=\textwidth]{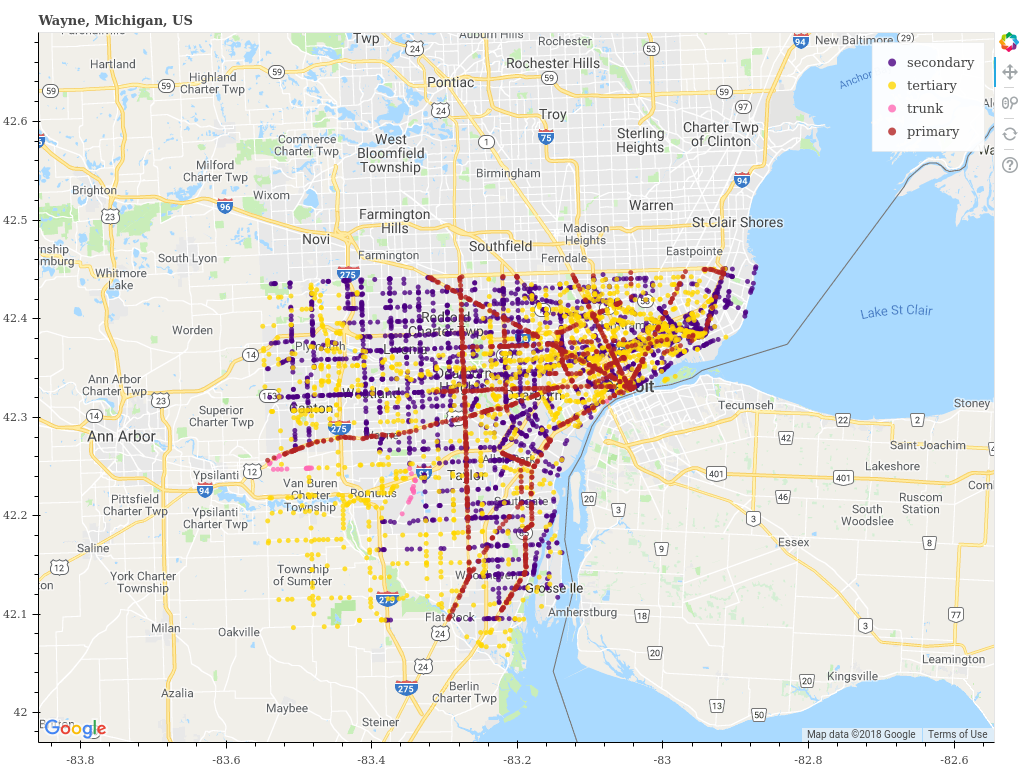}
\end{figure}

Next, we used the Google Street View API to download images at the starting point of each of the random road segment. Sometimes the Google API came back empty. We take the proportion of failed queries as an estimate of Google Street View coverage of the primary, secondary, tertiary, and trunk roads in the respective city. In Dhaka, for instance, just about 24.6\% of \href{https://github.com/geosensing/streetsense/scripts/google\_street\_view_Mturk-Dhaka.ipynb}{queries were successful}. (Figure \ref{fig:google_dhaka} plots the sampled locations.) Given the low coverage of Dhaka, we dropped Dhaka. In all, we have images of 978 locations for Bangkok, 872 for Jakarta, 999 for Lagos, and 4,828 for Wayne. Each photo captures a small segment of the road. (All the photos are available on \href{https://dataverse.harvard.edu/dataset.xhtml?persistentId=doi:10.7910/DVN/L3HN0K}{Harvard Dataverse}.)

Next, we recruited workers on Amazon's Mechanical Turk (MTurk) to code the images for the condition of the roads. To ensure quality, we only recruited `master' workers. We asked them if the segment of the road in the image had any 1) cracks, and 2) potholes. We also asked them, "if there are any road markings on the road, are they clear?" Lastly, we asked them, if there any litter and if the sidewalks were paved. The final survey for Bangkok, Jakarta, and Wayne, MI was the same (see ~\ref{fig:mturk_bangkok}).\footnote{We initially got Jakarta's images coded using alternate instrumentation (see \ref{fig:mturk_jakarta}). But we were concerned that this would lead to incommensurability. So we did another round of data collection with the same instrument.} Lagos' survey differed in very minor ways from Bangkok, Jakarta, and Wayne's (see ~\ref{fig:mturk_lagos}). We paid MTurkers 5 cents for answering the short survey for each image. To ensure quality, we also checked a few images at random to see if the coding was reasonable. We found one instance where one worker's judgments seemed really off and decided to reject those HITs.

\section*{Results}
Lest the readers miss an obvious point, before we present the results, we would like to draw their attention to it. Differences in the quality of roads across cities do not by default capture the extent of the road network. The extent of road network is easy to compute and regularly cited. Our contribution is  measurement of quality of roads, sidewalks, and litter on the streets efficiently.

The proportion of road segments with potholes is Jakarta is an astonishing .23. The commensurate number for Bangkok, Lagos, and Wayne is between .06--.07. But what does that mean? As we mentioned above, each image captures a small segment of the street. If we assume that a photo captures .5km, the expected number of potholes on a 10 km journey in Jakarta would be 2.3. That would make for a somewhat of a rough ride.

When it comes to cracks in the road, Wayne takes the top spot---the proportion of segments in Wayne with cracks is .62 followed by .44 for Jakarta and .20 and .24 for Bangkok and Lagos respectively. The high proportion is not particularly noteworthy for Wayne given its latitude, but it is noteworthy for Jakarta. 

Jakarta is also the dirtiest of the 4 cities with .21 of the segments containing litter.  Lagos comes second with .15 of the segments with litter. Lagos also takes the bottom spot for paved sidewalks---just .30 of the segments have a paved sidewalk.

\begin{table}[h]
\centering
\caption{Condition of the Roads in Different Places.}
\label{my-label}
\begin{tabular}{@{}lllllll@{}}
\toprule
city    & potholes & cracks & clear road markings & roads w/ markings & litter & paved sidewalk \\ \midrule
bangkok & .06      & .24    & .81                 & .98               & .06    & .51            \\
jakarta & .23      & .44    & .34                 & .97               & .21    & .49            \\
lagos   & .06      & .20    & .23                 & .95               & .15    & .30            \\
wayne   & .07      & .62    & .60                 & .90               & .09    & .67            \\ \bottomrule
\end{tabular}
\end{table}

Given there are differences across cities in the proportion of trunk, primary, secondary, and tertiary roads in the road network, we checked if cross-city comparisons are mostly capturing differences in road types than differences in conditions within each type of road. To examine this, we regressed the appropriate variable (whether or not there is a pothole, a crack) on the type of the road and city. Compared to tertiary roads, potholes are more common on primary roads (Diff. = .03), secondary roads (Diff. = .01), and trunk roads (Diff. = .05). But adjusting for the type of road doesn't change the across-city estimates much. For instance, the difference in the proportion of segments with potholes between Wayne and Jakarta is still .16.

Moving to cracks in the road, compared to tertiary roads, primary, secondary, and trunk roads have fewer cracks with differences being -.05, -.03, and -.09 respectively. Like with potholes, adjusting for the kind of roads doesn't seem to make much of a difference for inferences from raw data for cross-city comparison. 

Next, we analyzed the relationship between the condition of the roads and local income. To do that, we used the AskGeo API to get information on per capita income the census tract in which the lat/long lay. And we regressed whether a segment had a crack (or a pothole) or not on income split into quintiles. 

Before we present the results, a caveat. Given that we expect the largest relationship between quality of neighborhood roads and local income, we expect that our subsetting on primary, secondary, tertiary, and trunk roads to lead to smaller coefficients. 

Compared to road segments in tracts with per capita income less than 12k, the proportion of road segments with potholes in tracts with per capita income between 12k and 17k  was -.01 less. The proportion of road segments in tracts with per capita income of 17k to 23k  was -.02 less. For tracts with 23k and 29k, it was -.03 fewer segments with potholes, and for tracts with income between 29k and 83k, -.02 fewer segments had potholes. The relationship between local income and the proportion of segments with cracks was more uneven. The highest quintile had the fewest cracks but roads in the second and third income quintile areas had roughly the same number of cracks.

\section*{Discussion}

What is the condition of the streets? Are the streets paved? Do the streets have proper traffic signs and road markings? Is there litter on the streets? What proportion of vehicles on the streets is two-wheeled? And what proportion is man-powered, e.g., rickshaws? These are some of many the questions we can answer with \href{https://www.google.com/streetview/}{Google Street View}. In this paper, we provide a scalable way to answer such questions. We capitalize on \href{https://www.google.com/streetview/}{Google Street View}, pairing it with an open source Python package to randomly sample locations on the streets and crowdsourcing, to learn a host of compelling facts. 

The method that we describe here can be easily extended to automate the production of answers. Given that we are technically building a large labeled dataset, an obvious next step is to build a supervised machine learning infrastructure on top of it. Such an infrastructure can then provide automated estimates on many of these questions, along with useful caveats around coverage.

\clearpage
\bibliographystyle{apsr}
\bibliography{streets}
\clearpage
\appendix

\renewcommand{\thesection}{SI \arabic{section}}
\renewcommand\thetable{\thesection.\arabic{table}}  
\renewcommand\thefigure{\thesection.\arabic{figure}}
\counterwithin{figure}{section}

\section{Sampled Locations in Dhaka}
\begin{figure}[H]
\centering
\caption{Sampled Locations in Wayne} \label{fig:google_dhaka}
	\includegraphics[width=\textwidth]{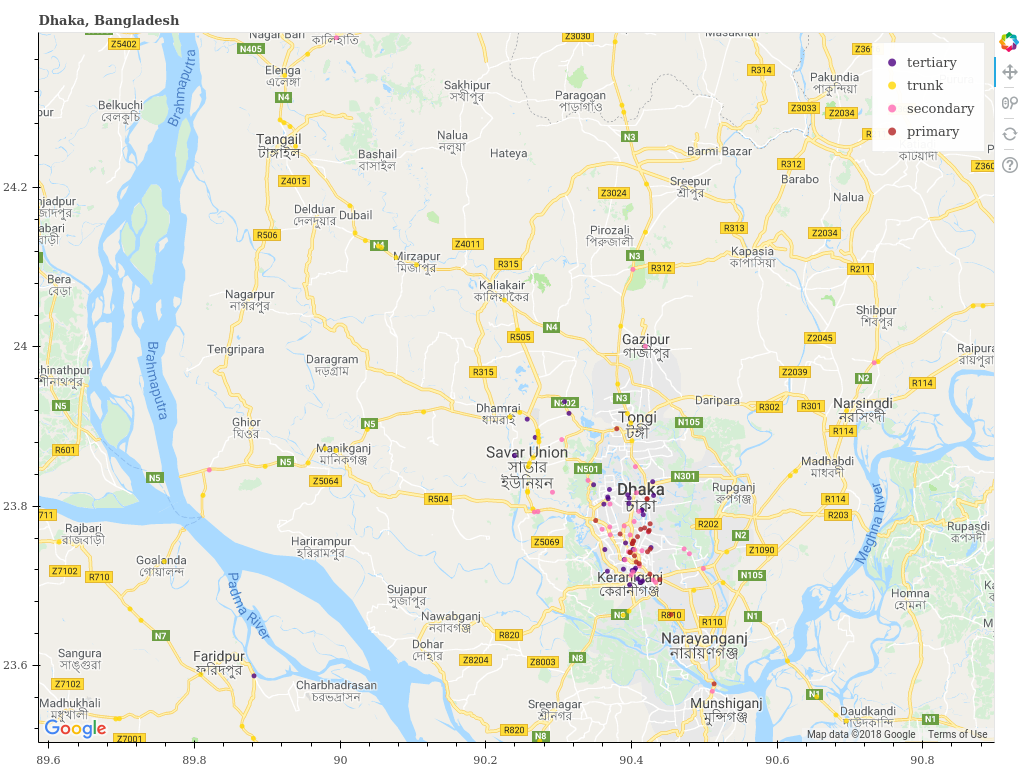}
\end{figure}
\clearpage
\section{Plots of Sampled Locations}
\begin{sidewaysfigure}
\centering
\setlength{\abovecaptionskip}{5pt plus 3pt minus 2pt}\caption{Sampled Locations in Bangkok}\label{fig:bangkok_sample}
\includegraphics[scale=0.5]{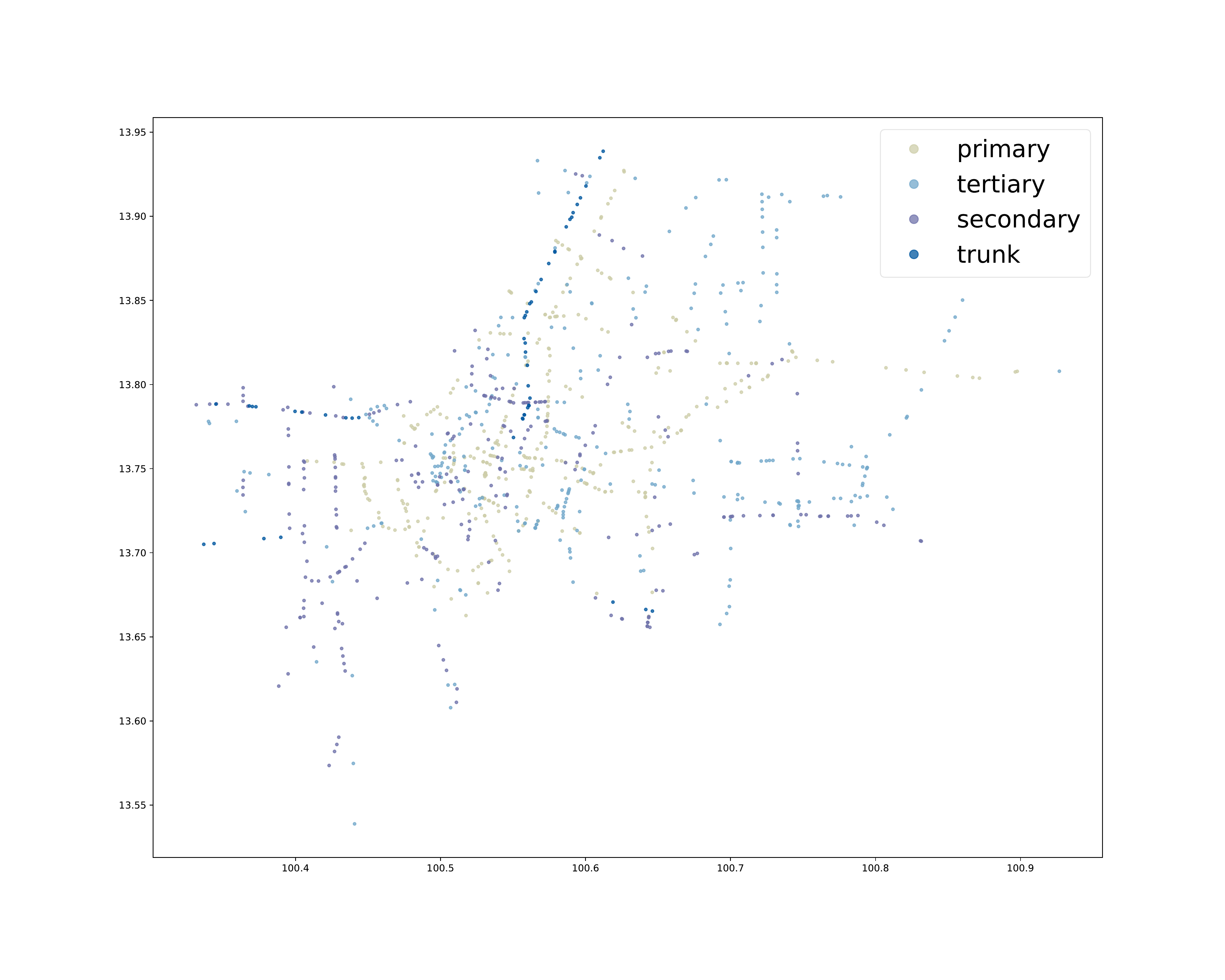}
\end{sidewaysfigure}

\begin{sidewaysfigure}
\centering
\caption{Sampled Locations in Jakarta}\label{fig:jakarta_sample}
\includegraphics[scale=0.5]{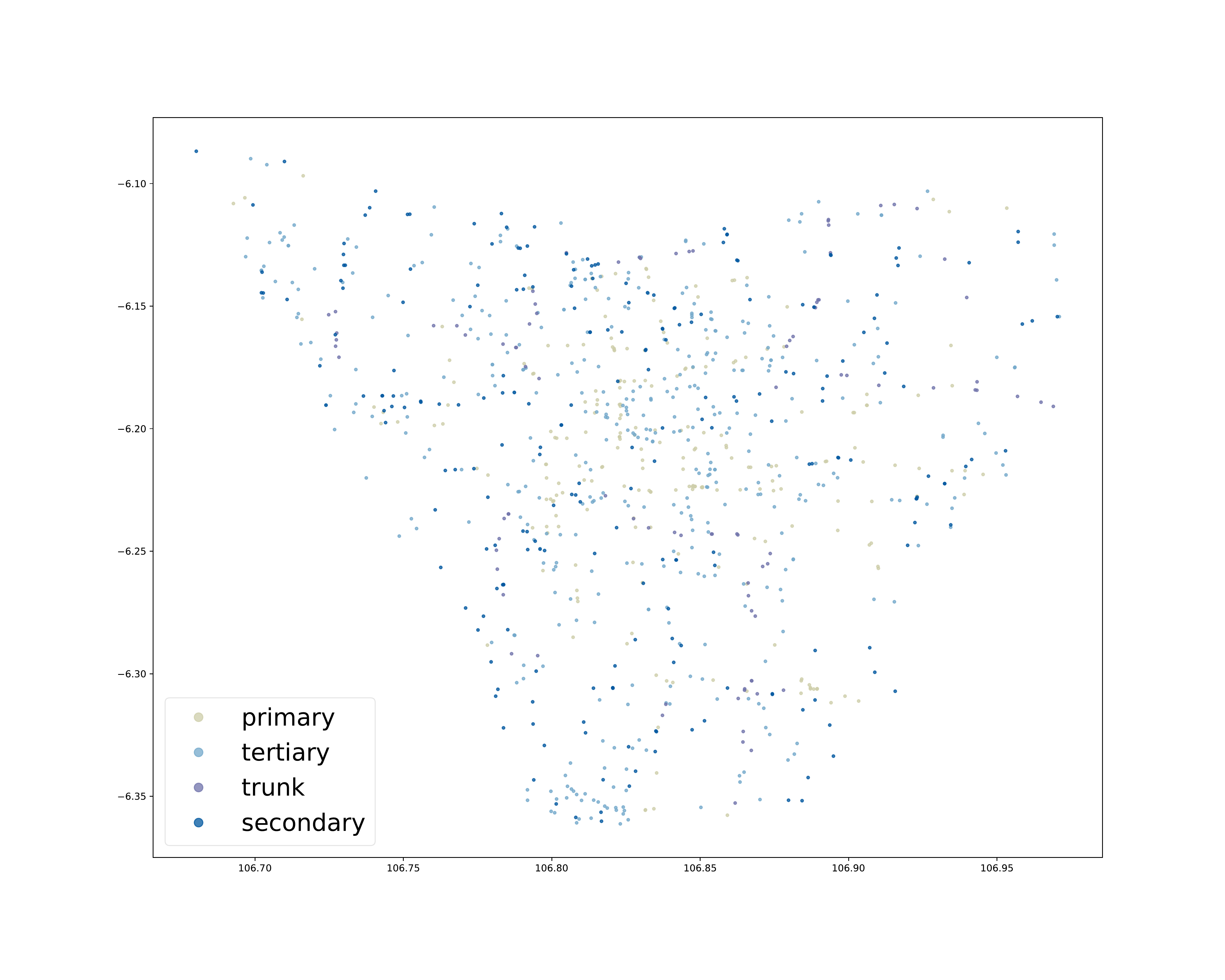}
\end{sidewaysfigure}

\begin{sidewaysfigure}
\centering
\caption{Sampled Locations in Lagos}\label{fig:lagos_sample}
\includegraphics[scale=0.5]{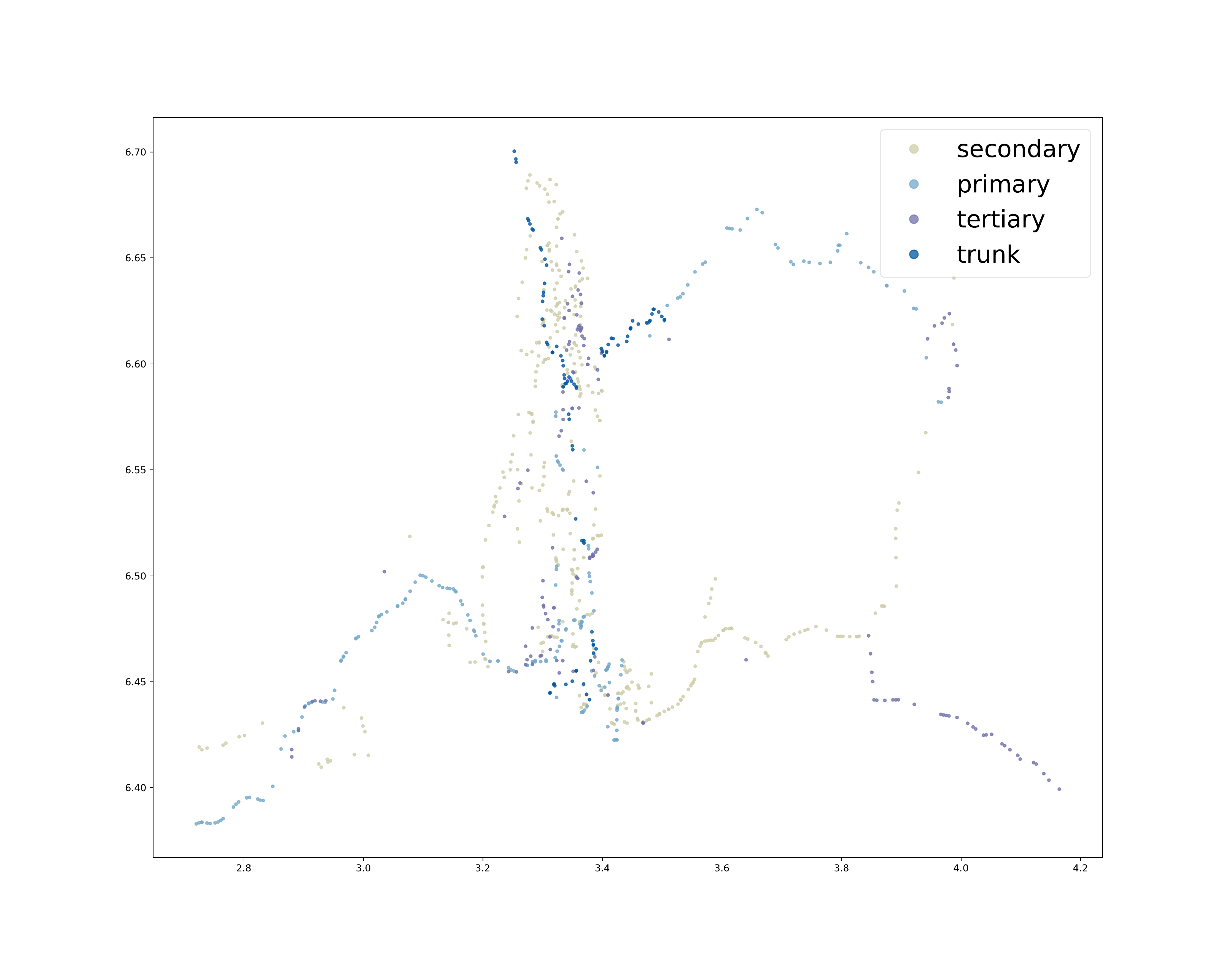}
\end{sidewaysfigure}

\begin{sidewaysfigure}
\centering
\caption{Sampled Locations in Wayne, MI}\label{fig:wayne2_sample}
\includegraphics[scale=0.5]{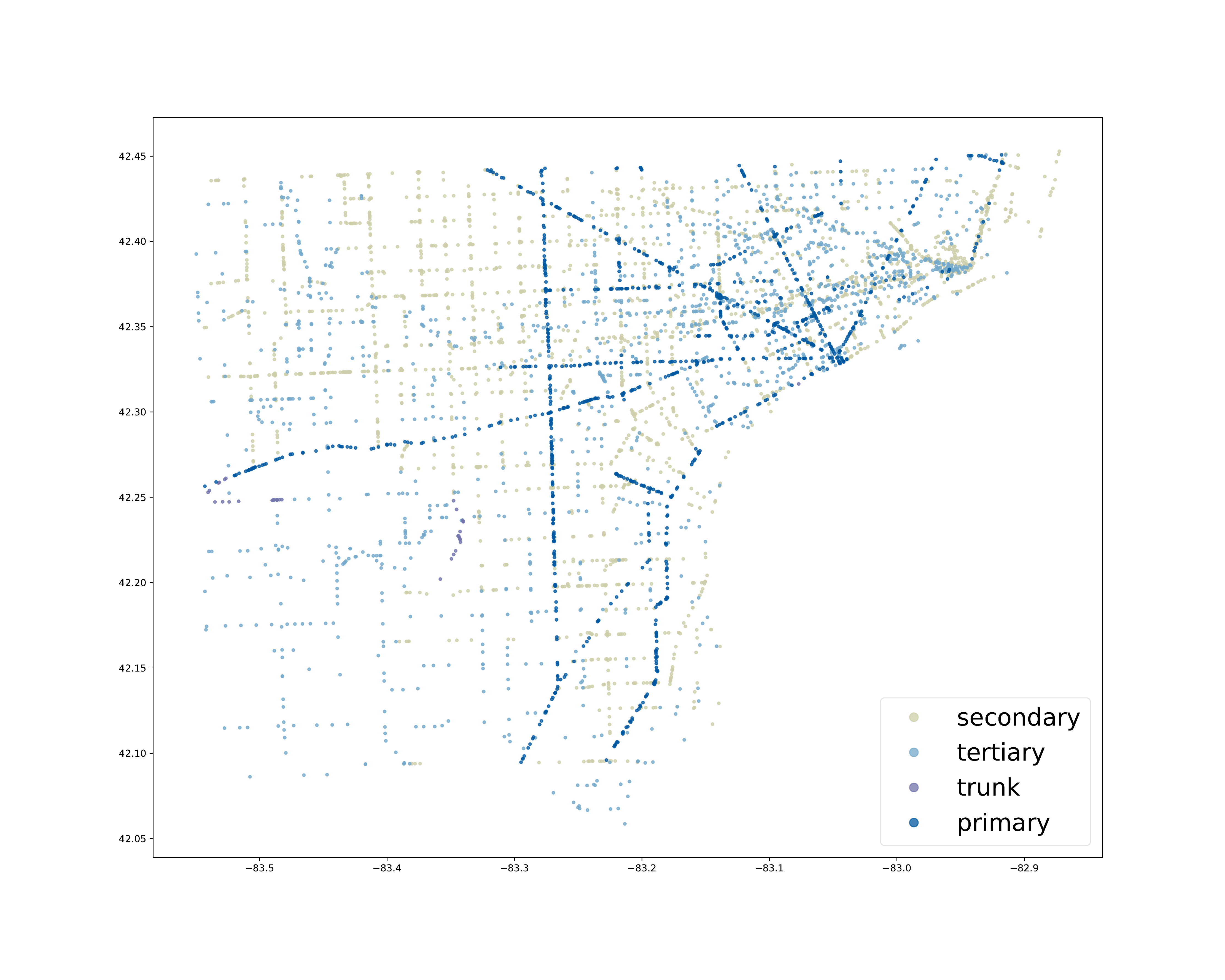}
\end{sidewaysfigure}

\clearpage
\section{Mturk Surveys}
\label{mturk_surveys}
\begin{figure}[h]
\centering
\caption{Screenshot of the Bangkok, Jakarta, and Wayne Questionnaire}\label{fig:mturk_bangkok}
\includegraphics[scale=0.5]{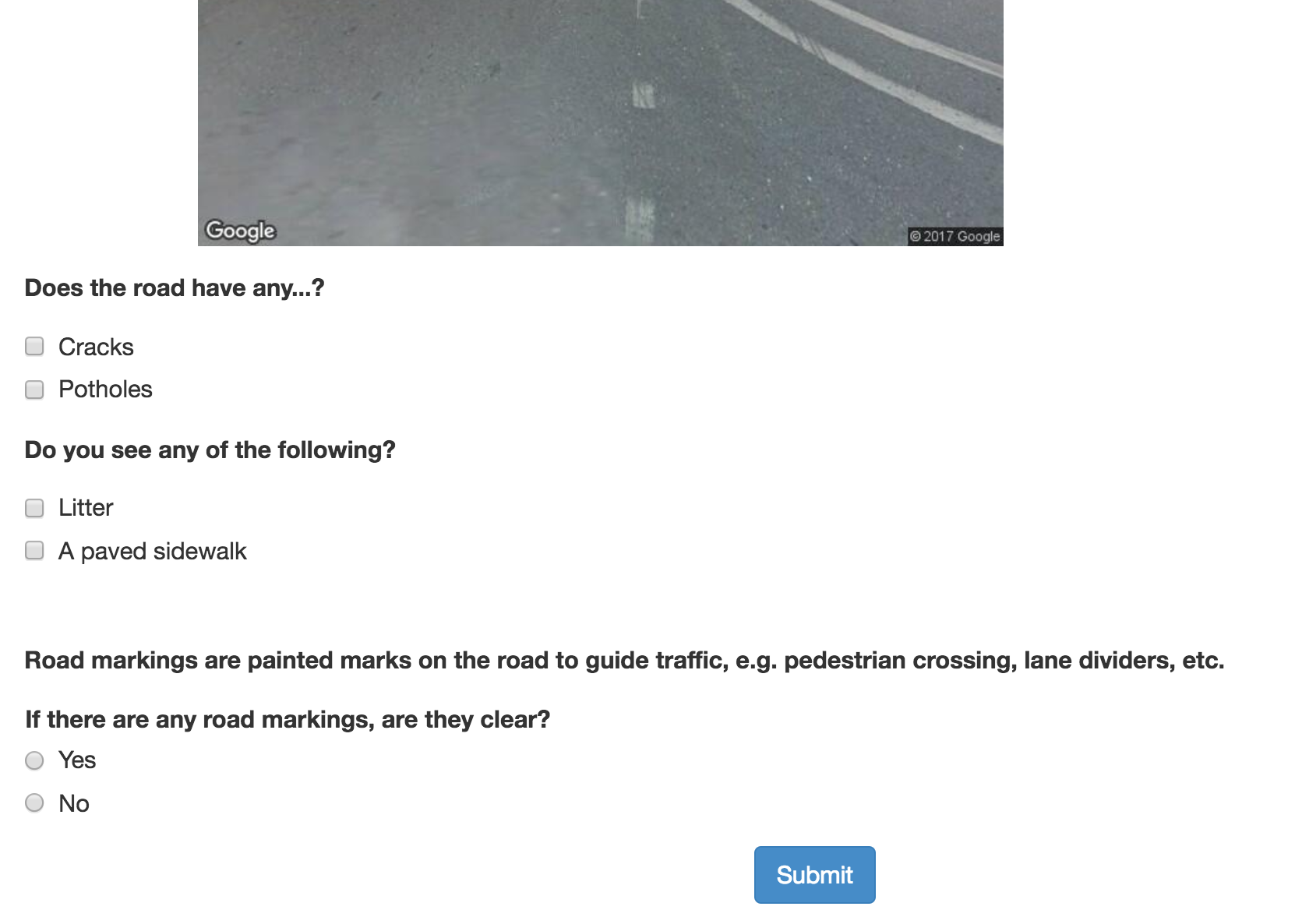}
\end{figure}

\clearpage

\begin{figure}
\centering
\caption{Screenshot of the Lagos Questionnaire}\label{fig:mturk_lagos}
\includegraphics[scale=0.5]{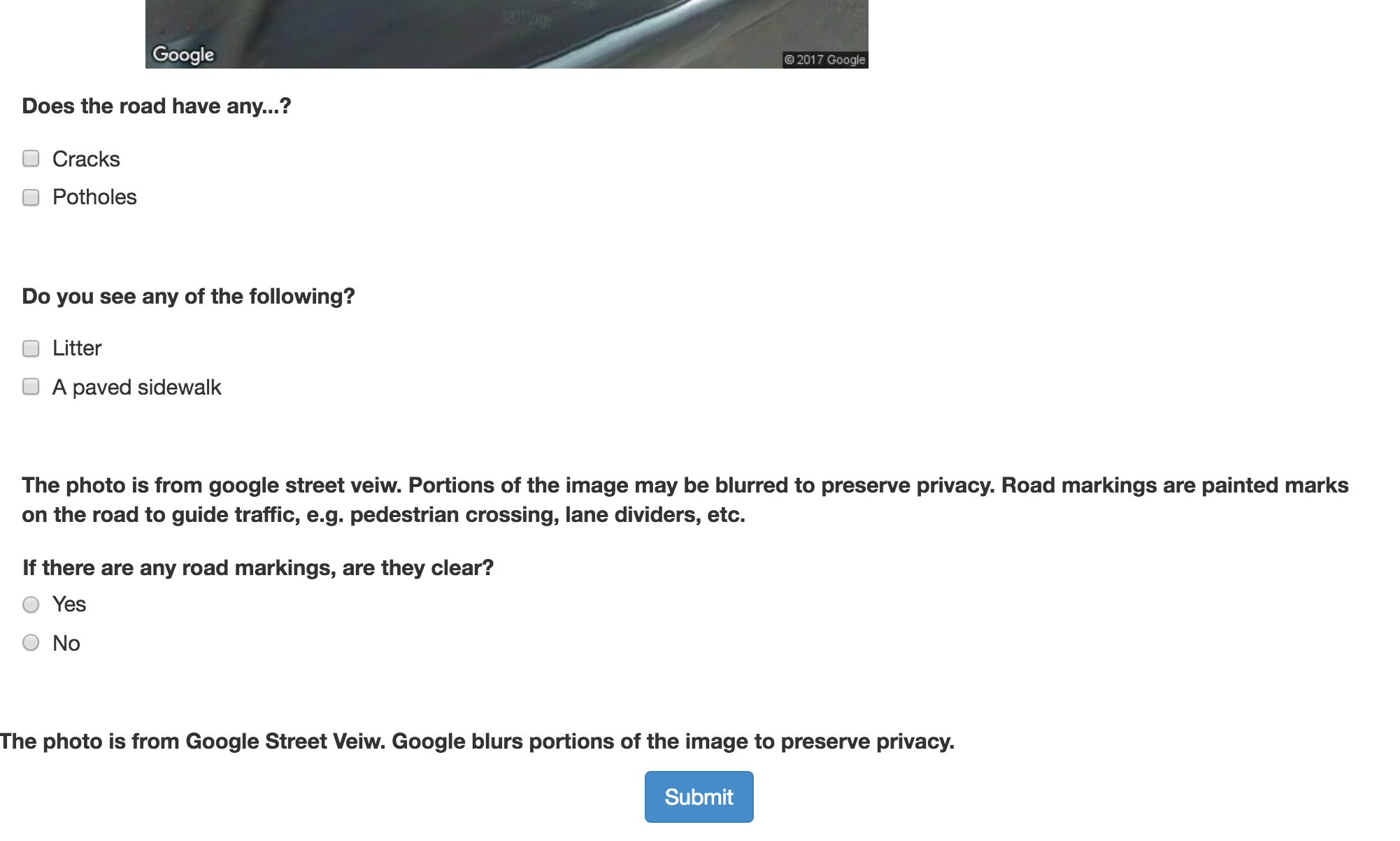}
\end{figure}

\clearpage

\begin{figure}
\centering
\caption{Screenshot of the First Jakarta Questionnaire}\label{fig:mturk_jakarta}
\includegraphics[scale=0.5]{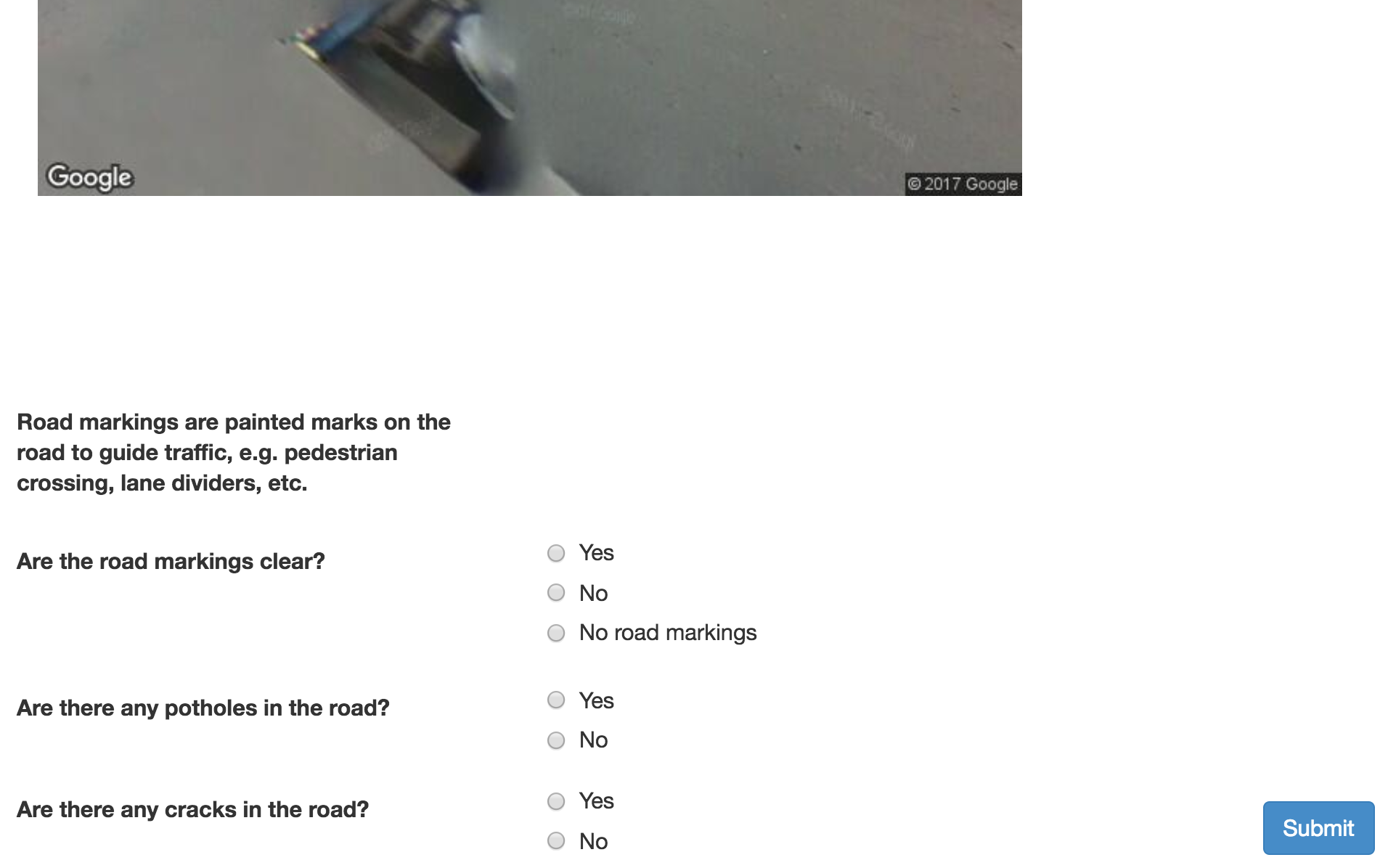}
\end{figure}

\end{document}